\documentclass[fleqn,usenatbib]{mnras}

\usepackage{newtxtext,newtxmath}
\usepackage[T1]{fontenc}
\usepackage{ulem}

\DeclareRobustCommand{\VAN}[3]{#2}
\let\VANthebibliography\thebibliography
\def\thebibliography{\DeclareRobustCommand{\VAN}[3]{##3}\VANthebibliography}

\usepackage{graphicx}	% Including figure files
\usepackage{amsmath}	% Advanced maths commands
\usepackage{soul}
\usepackage{xcolor}

\newcommand{\fornax}{F\textsc{ornax}}

\title[Three approaches for classification of PNS modes]{Three approaches for the classification of protoneutron star oscillation modes}

\author[M. C. Rodriguez et al.]{
M. C. Rodriguez,$^{1,2}$\thanks{E-mail: mcrodriguez@fcaglp.unlp.edu.ar}
Ignacio F. Ranea-Sandoval,$^{1,2}$
C. Chirenti$^{3,4,5,6}$,
and D. Radice$^{7,8,9}$\thanks{Alfred P. Sloan fellow}
\\
$^{1}$Grupo de Gravitación, Astrofísica y Cosmología, Facultad de Ciencias Astronómicas y Geofísicas, Universidad Nacional de La Plata,
\\Paseo del Bosque S/N, 1900, La Plata, Argentina.\\
$^{2}$CONICET, Godoy Cruz 2290, 1425, CABA, Argentina.\\
$^{3}$Department of Astronomy, University of Maryland, College Park, MD 20742-2421, USA.\\
$^{4}$Astroparticle Physics Laboratory NASA/GSFC, Greenbelt, MD 20771, USA.\\
$^{5}$Center for Research and Exploration in Space Science and Technology, NASA/GSFC, Greenbelt, MD 20771, USA.\\
$^{6}$Center for Mathematics, Computation, and Cognition, UFABC, Santo André, SP 09210-170, Brazil.\\
$^{7}$Institute for Gravitation and the Cosmos, The Pennsylvania State University, University Park, PA 16802, USA.\\
$^{8}$Department of Physics, The Pennsylvania State University, University Park, PA 16802, USA.\\
$^{9}$Department of Astronomy $\&$ Astrophysics, The Pennsylvania State University, University Park, PA 16802, USA.\\
}

\date{Accepted XXX. Received YYY; in original form ZZZ}

\pubyear{2023}

\begin{document}
\label{firstpage}
\pagerange{\pageref{firstpage}--\pageref{lastpage}}
\maketitle

\begin{abstract}
The future detection of gravitational waves (GWs) from a galactic core-collapse supernova will provide information on the physics inside protoneutron stars (PNS). In this work, we apply three different classification methods for the PNS non-radial oscillation modes: Cowling classification, Generalized Cowling Nomenclature (GCN), and a Classification Based on Modal Properties (CBMP). Using PNS models from $3$D simulations of core-collapse supernovae, we find that in the early stages of the PNS evolution, typically before $0.4$ seconds after the bounce, the Cowling classification is inconsistent, but the GCN and the CBMP provide complementary information that helps to understand the evolution of the modes. In the GCN, we note several avoided crossings as the mode frequencies evolve at early times, while the CBMP tracks the modes across the avoided crossings. We verify that the strongest emission of GWs by the PNS corresponds to the $f$-mode in the GCN, indicating that the mode trapping region alternates between the core and the envelope at each avoided crossing. At later times, approximately $0.4$ seconds after the bounce, the three classification methods present a similar description of the mode spectrum. We use our results to test universal relations for the PNS modes according to their classification and find that the behaviour of the universal relations for $f$- and $p$-modes is remarkably simple in the CBMP. 
%It should be a single paragraph not more than 250 words (200 words for Letters). No references should appear in the abstract.
\end{abstract}


%%%%%%%%%%%%%%%%% BODY OF PAPER %%%%%%%%%%%%%%%%%%%%%%%%%%%%%%%%%%%%%%%%%%%%%%
\section{Introduction}

In recent years, the first detection of gravitational waves (GWs) emitted by the merger of binary systems of black holes (BHs) \citep{Abbott2016} and neutron stars (NSs) \citep{Abbott2017b} has given rise to a new field of research that could offer information about the physics of these compact objects. The double NS coalescence GW170817 has also been detected through the entire electromagnetic spectrum and thus added GWs to the landscape of multimessenger astronomy. These events, together with the astronomical properties of the $\sim 100$ mergers of compact binaries detected so far \citep{Abbott2021}, help to constrain the properties of BHs and NSs. The GWTC-2 and GWTC-3 catalogues contain the properties inferred from the detection of GWs \citep{abbott2021a,abbott2021gwtc3}. 

A complementary view of the physics of compact objects will emerge upon the detection of GWs from a Galactic core-collapse supernova. One of the possible remnants of a core-collapse supernova is a protoneutron star (PNS) \citep[see, for example, ][and references therein]{burrows:2021ccs}. The resulting GW emission is associated with PNS oscillations and dynamics  \citep{Murphy2009,Marek2009, Muller2013}. Unlike NS binary coalescences, for which the stars are effectively cold, inside a PNS temperatures are initially above $5 \times 10^{11}$ K. Over a timescale of a few seconds, PNSs become more compact and cool down until they reach an equilibrium in which they are considered NSs \citep{Pons1999}. During this time, the microphysics inside the star also changes due to processes such as electron capture and, later, neutrino emission. As a PNS cools, its central densities rapidly become higher than nuclear saturation density. At such densities, the equation of state (EoS), which gives us information about the relation between the pressure and the energy density inside the star, is still largely unknown. 

The study of the GW signal from simulations of core-collapse supernovae has been performed in two dimensions with axial symmetry, e.g., \cite{Marek2009,Murphy2009,Cedra-Duran2013,Muller2013,Abdikamalov2014,Yakunin2015, Morozova2018,Pan2018}, among others, and, in the last years, in three dimensions \citep[see, for example,][]{Kuroda2016,Kuroda2017,Hayama2016,Andersen2017,Hayama2018,Oconnor2018,Powell2019,Mezzacappa2020,Vartanyan2022,Nakamura2022,Powell2022,mezzacappa:2023ccs,Vartanyan:2023sxm}, in which the treatment is much more precise and thus computationally more expensive than in two dimensions. 

The complexity of the processes during core collapse renders it impossible to predict with high precision the resulting gravitational waveforms.  Nonetheless, the community has reached some agreement on the key aspects of the signal morphology after several years of numerical simulations \citep{Mezzacappa2020b,Oconnor2018b}. Therefore, a future detection of GWs from Galactic core-collapse supernovae will add more information about the physics inside the PNS and will facilitate tests of the simulated models \citep{Abdikamalov2022,Wolfe2023}. 

However, detecting the emitted GWs produced in core-collapse supernovae will be challenging. Knowledge of the morphology of the emitted waveform is of paramount importance for developing effective algorithms capable of extracting GW data from the detected signal. Different methods have been proposed: examples include principal component analysis \citep{Heng2009,Rover2009,Powell2017,Roma2019}, inference using Bayesian analysis \citep{Rover2009,Powell2017,Roma2019,Summerscales2008, Gill2018}, denoising techniques \citep{Mukherjee2017}, and machine learning \citep{Cavaglia2020,Lopez2020,Iess2020,Chan2020,Astone2018}.  One of the most promising ways to detect GWs from nearby core-collapse supernovae uses the excess-power coherent waveburst search, which has already detected GWs produced in mergers of compact objects \citep{Abbott2019,abbott2021a,abbott2021gwtc3}, even in low-latency (e.g., for GW150914 \citep{Abbot2017}).

Here, however, we focus on a different approach, using a theoretical framework which has been developed to study stellar perturbations.  We use this framework to describe the GW signal of a supernova as the superposition of the oscillation modes of the PNS  \citep{Murphy2009,Marek2009,Muller2013}. Oscillation modes of stellar configurations can be classified by taking into account their main restoring force. Generally speaking, the dominant restoring force is pressure for $p$-modes (and $f$-modes) and buoyancy for $g$-modes \citep{Cowling1941,Kokkotas1999}. In the standard picture, two local quantities determine the character of a  mode with given frequency $\sigma/(2\pi)$: the Lamb frequency, $\mathcal{L}$, and the Br\"unt-V\"ais\"al\"a frequency, $\mathcal{N}$. Sound waves can propagate in regions of the star in which $\sigma^2$ is greater than both $\mathcal{L}^2$ and $\mathcal{N}^2$, while gravity waves are possible in regions of the star in which $\sigma ^2$ is smaller than both $\mathcal{L}^2$ and $\mathcal{N} ^2$ \citep[see, for example, ][for a review]{unno1989nonradial}.

Usually, the classification of non-radial oscillation modes is based on the classification scheme proposed by \citet{Cowling1941}, in which the modes are defined by their restoring force and the number of nodes in the radial eigenfunction. But, for stars in which the stellar dynamics is different from main sequence stars (or cold NSs), such as evolved stars and newly born PNS, the classification is not so trivial and the Cowling classification is not enough to correctly identify the non-radial oscillation modes. 
For this reason, in section \ref{sec:classification} we explore alternative classification methods of PNS modes, which were developed in the context of the study of non-radial pulsations of evolved stars and give complementary information about the modes, enabling a proper classification in this setting.  

Following \citet{unno1989nonradial}, we use the Generalized Cowling Nomenclature (GCN; \citep{Eckart1960,Scuflaire1974,Osaki1975}) and the Classification Based on Modal Properties (CBMP) of \citet{Shibahashi1976}. The GCN was also applied to PNS simulations by \citet{Torres2019} and other subsequent works. The CBMP provides qualitatively similar results to the matching classification based on the shape of the radial eigenfunction proposed by \citet{Torres2019} and to the ramp-up mode description discussed by \citet{Mori2023}, but the CBMP is based on a more rigorous evaluation of the kinetic energy of the mode in its trapped region. These complementary classification schemes help us to understand and identify different features of the modes, especially in the early stages of the PNS after the bounce. 

The aim of our work is to be able to identify correctly the non-radial oscillations modes of PNSs. By using complementary classification schemes, we are able to capture information from GW signals that can help to infer properties of the pulsating PNS. To obtain the modes, we apply linear perturbation theory to the PNS and use a Lagrangian description of the fluid perturbation, following \cite{Morozova2018}. 

In section \ref{sec:2} we present the setup for calculating the non-radial oscillation modes from PNS described by data from core-collapse supernova simulations \citep{Radice2019}. In section \ref{sec:classification}, we describe the three different mode classification schemes.  In section \ref{sec:results}, we present the results obtained using the three different classification methods and a brief analysis of the universal relations of PNSs. Finally, in section \ref{sec:conclusions} we summaryze our work and present our conclusions.

%%%%%%%%%%%%%%%%%%%%%%%%%%%%%%%%%%%%%%%%%%%%%%%%%%%%%%%%%%%%%%%%%%%%%%%%%%%%%%%%
\section{Protoneutron star non-radial oscillations}
\label{sec:2}

We use hydrostatic PNS configurations obtained from 3D core-collapse supernovae simulations of \cite{Radice2019} following the approach of \citet{Sotani2016}. Namely, the results of the 3D simulations are averaged to construct an effective barotropic equation of state, which we use to solve the Tolman–Oppenheimer–Volkoff equations to reconstruct the structure of the PNS and the spacetime geometry. See also \citet{Westernacher-Schneider:2020bkw} for an alternative approach that is fully consistent with the pseudo-Newtonian formalism employed in the dynamical simulations. The line element that describes a static spherically symmetric conformally flat spacetime in isotropic coordinates ($t$, $x_i$) is given by
\begin{equation}
    ds^2=g_{\mu \nu} dx^{\mu}dx^{\nu}=- \alpha^2 dt^2 + \psi^4 f_{ij} dx^i dx^j,
\end{equation}
where $\alpha$ is the lapse function, $\psi$ is the conformal factor whose value is set to $1$, and $f_{ij}$ is the flat spatial 3-metric. The conformal flatness approximation incorporates the spacetime dynamics evolution, but it ignores gravitational radiation \citep{Isenberg2008, Wilson1989}. 

To describe the fluid perturbations we used the Lagrangian displacement vector
\begin{equation}
    \xi^i=(\xi^r, \xi^{\theta}, \xi^{\phi}),
\end{equation}
with
\begin{eqnarray}
    \xi^r&=&\eta_r Y_{lm} e^{-i\sigma t}, \nonumber \\
    \xi^{\theta}&=&\eta_{\perp} \frac{1}{r^2} \partial _\theta Y_{lm} e^{-i\sigma t}, \\
    \xi^{\phi}&=&\eta_{\perp} \frac{1}{r^2 \sin^2 \theta}  \partial _\phi Y_{lm} e^{-i\sigma t}, \nonumber
\end{eqnarray}
where $Y_{lm}$ are the spherical harmonics and the functions $\eta_r$ and $\eta_{\perp}$ depend only on the radial coordinate $r$. For each of the perturbing functions, a time dependence $e^{i\sigma t}$ is assumed.

In order to find the quasinormal modes, we need to solve a set of four first-order coupled differential equations for $\eta_r$, $\eta _\perp$ and two additional functions $f_\alpha$ and $\delta \hat{\alpha}$, with appropriate values at the centre and a boundary condition, as done in \cite{Morozova2018}. The scalar function $\delta \hat \alpha$ is the amplitude of the lapse function perturbation that depends only on the radial coordinate, i.e. $\delta \alpha = \delta \hat \alpha(r) Y_{lm} e^{-i\sigma t}$ and $f_{\alpha}$ is defined as $f_{\alpha} = \partial_r (\delta \hat \alpha / \alpha)$. The equations for the four unknown functions $\eta_r$, $\eta _\perp$, $f_\alpha$ and $\delta \hat{\alpha}$ are
\begin{eqnarray} 
0&=&\partial_{r} \eta_{r}+\left[\frac{2}{r}+\frac{1}{\Gamma_{1}} \frac{\partial_{r} P}{P}+6 \frac{\partial_{r} \psi}{\psi}\right] \eta_{r} \nonumber \\ \label{eqn:1}
&&+\frac{\psi^{4}}{\alpha^{2} c_{s}^{2}}\left(\sigma^{2}-\mathcal{L}^{2}\right) \eta_{\perp}
-\frac{1}{\alpha c_{s}^{2}} \delta \hat{\alpha}, \\
0&=&\partial_{r} \eta_{\perp}-\left(1-\frac{\mathcal{N}^{2}}{\sigma^{2}}\right) \eta_{r}+\left[\partial_{r} \ln q-\tilde{G}\left(1+\frac{1}{c_{s}^{2}}\right)\right] \eta_{\perp} \nonumber \\ \label{eqn:2}
&&-\frac{1}{\alpha \tilde{G}} \frac{\mathcal{N}^{2}}{\sigma^{2}} \delta \hat{\alpha}, \\
0&=&\partial_{r} f_{\alpha}+\frac{2}{r} f_{\alpha}+4 \pi\left[\partial_{r} \rho-\frac{\rho}{P \Gamma_{1}} \partial_{r} P\right] \eta_{r} \nonumber\\ \label{eqn:3}
&&-\frac{4 \pi \rho}{P \Gamma_{1}} q \sigma^{2} \eta_{\perp}
+\left[\frac{4 \pi \rho^{2} h}{P \Gamma_{1} \alpha}-\frac{1}{\alpha} \frac{l(l+1)}{r^{2}}\right] \delta \hat{\alpha},\\ 
0&=&\partial_{r} \delta \hat{\alpha} - f_{\alpha} \alpha+\tilde{G} \delta \hat{\alpha}, \label{eqn:4}
\end{eqnarray}
where $\Gamma_1$ is the adiabatic index, $P$ is the pressure, $\rho$ is the mass density, $h$ is the specific enthalpy, $c_s$ is the speed of sound and $q = \rho h \alpha^{-2} \psi^4$; all of these quantities refer to the interior of the PNS. The Brunt-V\"ais\"ala and Lamb frequencies are 
\begin{eqnarray}
    \mathcal{N}^2&=&\frac{\alpha \partial_r \alpha}{\psi^4} \left( \frac{1}{\Gamma_1}\frac{\partial_r P}{P} - \frac{\partial_r e}{\rho h} \right), \\
    \mathcal{L}^2&=&\frac{\alpha^2}{\psi^4}c_s^2 \frac{l(l+1)}{r^2}.
\end{eqnarray}
where $e$ is the energy density. The radial component of the gravitational acceleration $\tilde{G}$ is defined as $\tilde{G} =\frac{1}{\rho h}\partial_r P$.\footnote{An alternative definition is given by $\tilde{G} = -\partial_r \ln \alpha$. Both definitions of $\tilde {G}$ were tested in our code giving very similar results as  expected.} 

To solve equations (\ref{eqn:1})-(\ref{eqn:4}), the boundary condition $\Delta P = 0$ on the Lagrangian perturbation of the pressure has to be satisfied at the stellar radius, $R$. Following \cite{Morozova2018}, we define the stellar radius where $\rho = 10^{10} \text{g cm}^{-3}$. (Different values of the mass density for the stellar radius do not have a large impact on the mode frequencies; g-modes are essentially unchanged.) The  boundary condition is given by:
\begin{eqnarray}
    q\sigma^2 \eta_{\perp}-\frac{\rho h}{\alpha}\delta \hat \alpha + \partial_r P \eta_r =0. \label{eqn:bc}
\end{eqnarray} 
From eqs. (\ref{eqn:1})-(\ref{eqn:4}), regularity of the solutions at $r=0$ requires
\begin{eqnarray}
    \eta_r|_{r=0} = f_{\alpha}|_{r=0} &\sim& r^{l-1}, \label{eqn:bc1} \\
    \delta \hat \alpha|_{r=0} = \alpha(0) \eta_{\perp}|_{r=0} &\sim& \frac{\alpha(0)}{l} r^l. \label{eqn:bc2}
\end{eqnarray}
We note that \cite{Morozova2018} set the boundary conditions for $\delta \hat \alpha$ and $f_{\alpha}$   to zero. However, the difference in the resulting mode frequencies is negligible. 

We computed the oscillation frequencies and their respective eigenfunctions in \textsc{Fortran90}. Our code uses a shooting method with Ridders' root-finding algorithm to look for eigenfrequencies in the range from $100$ Hz to $2000$ Hz until the boundary condition is satisfied, solving eqs.  (\ref{eqn:1})-(\ref{eqn:4}) (or eqs. (\ref{eqn:5}) and (\ref{eqn:6}) below, in the Cowling approximation) with a Runge-Kutta-Fehlberg integrator. This is done at every time step of the evolution of the PNS, as described in Section \ref{sec:data}.

\subsection{Data} 
\label{sec:data} 

The profiles of evolving PNS were calculated from the $3$D simulations of core-collapse supernovae performed by \citet{Radice2019}. The simulations were performed using the {\fornax} code \citep{Skinner2019}, which solves the equations of neutrino-radiation-hydrodynamics using a sophisticated, multi-dimensional transport scheme. Gravity was handled using an effective general-relativistic potential following \citep{Marek2009}. The neutrino treatment accounted
for gravitational redshift, Doppler effects, and inelastic
scattering~\citep{Burrows2018}. NS matter was treated using the SFHo equation of state of \citet{Steiner2013}. In this work, we use models with $9 M_{\odot}$, $10 M_{\odot}$, $11 M_{\odot}$, $12 M_{\odot}$, $13 M_{\odot}$, $19 M_{\odot}$, $25 M_{\odot}$ and $60 M_{\odot}$ ZAMS mass. In table \ref{tab:models} we list the details of each model. For more details, see \citet{Radice2019} and \citet{Burrows2020}. 
\begin{table}
\begin{center}
\begin{tabular}{||c c c c||} 
 \hline
 Model & ZAMS mass & explode & Symbol \\ [0.5ex] 
 \hline\hline
 1 & $9 M_{\odot}$ & yes & $\blacktriangle$ \\
 \hline
 2 & $10 M_{\odot}$ & yes & $\bullet$ \\
 \hline
 3 & $11 M_{\odot}$ & yes & $\circ$ \\
 \hline
 4 & $11 M_{\odot}$ & yes & $\blacksquare$ \\
 \hline
 5 & $12 M_{\odot}$ & yes & $\square$ \\ [1ex]
 \hline
 6 & $13 M_{\odot}$ & no & $\ast$ \\ [1ex] 
 \hline
 7 & $19 M_{\odot}$ & yes & $\times$ \\ [1ex] 
 \hline
 8 & $19 M_{\odot}$ & yes & $+$ \\ [1ex] 
 \hline
 9 & $25 M_{\odot}$ & yes & $\bigtriangledown$ \\ [1ex] 
 \hline
 10 & $60 M_{\odot}$ & yes & $\bigtriangleup$ \\ [1ex] 
 \hline
\end{tabular}
\caption{3D simulations of core-collapse supernovae from \citet{Radice2019} corresponding to the PNS evolution models used in this work. Models 3 and 7 ignore the virial correction to the neutrino--nucleon scattering from \citet{Horowitz2016}, which is used in all other models.}
\label{tab:models}
\end{center}
\end{table}

\subsection{The relativistic Cowling approximation} \label{cowling}

As the classification methods described in section \ref{sec:classification} below are strictly valid only when the metric perturbations are neglected, we need to use the relativistic Cowling approximation \citep{mcdermott1983}, in which only fluid perturbations are considered. 
Under this approximation, the set of differential equations (\ref{eqn:1})-(\ref{eqn:4}) becomes
\begin{eqnarray} 
0&=&\partial_{r} \eta_{r}+\left[\frac{2}{r}+\frac{1}{\Gamma_{1}} \frac{\partial_{r} P}{P}\right] \eta_{r}+\frac{1}{\alpha^{2} c_{s}^{2}}\left(\sigma^{2}-\mathcal{L}^{2}\right) \eta_{\perp}, \label{eqn:5} \\
0&=&\partial_{r} \eta_{\perp}-\left(1-\frac{\mathcal{N}^{2}}{\sigma^{2}}\right) \eta_{r}+\left[\partial_{r} \ln q-\tilde{G}\left(1+\frac{1}{c_{s}^{2}}\right)\right] \eta_{\perp}. \label{eqn:6} 
\end{eqnarray}
In the relativistic Cowling approximation, these are the only equations needed to find the mode frequency, $\sigma$, whose associated eigenfunctions are $\eta_{r}$ and $\eta_{\perp}$. 

Boundary conditions need to be imposed to solve equations (\ref{eqn:5}$)-($\ref{eqn:6}). Under this approximation, the values of  $\sigma^2$ we are looking for will be the ones that satisfy
\begin{eqnarray}
    q\sigma^2 \eta_{\perp} + \partial_r P \eta_r =0. \label{eqn:bc_new}
\end{eqnarray} 
%%%%%%%%%%%%%%%%%%%%%%%%%%%%%%%%%%%%%%%%%%%%%%%%%%%%%%%%%%%%%%%%%%%%%%%%%%%%%%%%
\section{Classification of modes}
\label{sec:classification}

\begin{figure}
    \centering
    \includegraphics[width=.97\linewidth]{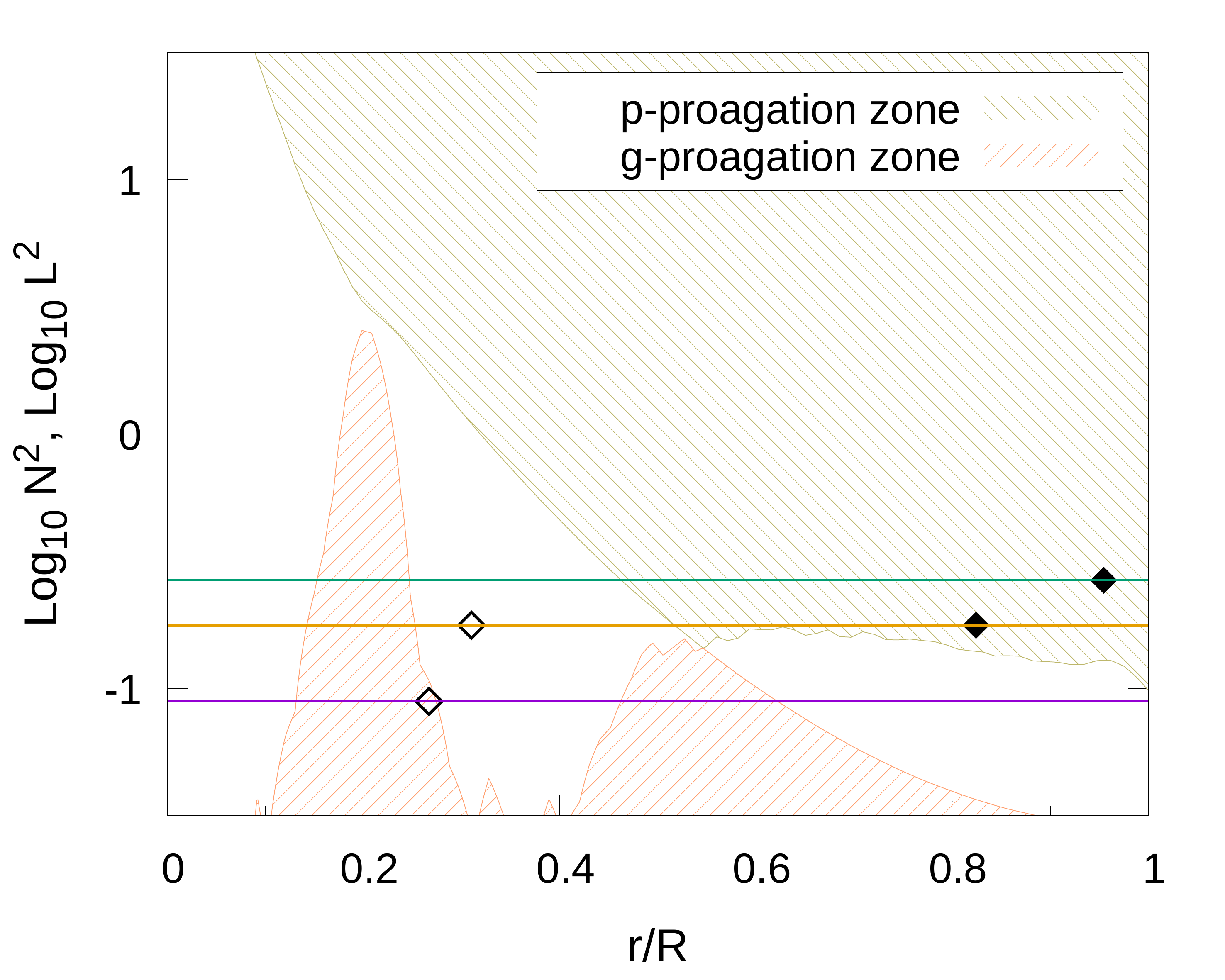}
     \caption{Propagation diagram for a PNS at $0.4$ s after the bounce, produced by the model with $9M_{\odot}$ progenitor mass (model 1 in Table \ref{tab:models}), for $l = 2$. The region above $\log_{10}(\mathcal{N}^2)$ and $\log_{10}(\mathcal{L}^2)$ is the p-propagation zone and the region below $\log_{10}(\mathcal{N}^2)$ and $\log_{10}(\mathcal{N}^2)$ is the g-propagation zone. Both $\mathcal{N}$ and $\mathcal{L}$ are normalized by $\sqrt{GM/R^3}$. The horizontal lines show the frequencies of three different modes; diamonds indicate the position of the nodes in the radial eigenfunction. Filled (empty) diamonds correspond to $p$-($g$-)nodes, see Section \ref{subsec:general_class}. 
     }
    \label{fig:prop_diagram}
\end{figure}

The analysis of the stellar oscillations gives information about the properties of the modes, which can be inferred in a propagation diagram like the one shown in figure~\ref{fig:prop_diagram}. This propagation diagram shows, for a PNS model, the  normalized Brunt-V\"ais\"ala and the Lamb frequencies as functions of the normalized radius of the star, and corresponds to perturbation modes with $l=2$.  In this diagram, the $p$-propagation zone is defined as the region above both $\log_{10}(\mathcal{N}^2)$ and $\log_{10}(\mathcal{L}^2)$ (represented with khaki colour pattern) and the $g$-propagation zone is defined as the region below both $\log_{10}(\mathcal{N}^2)$ and $\log_{10}(\mathcal{L}^2)$ (represented with orange colour pattern). Frequencies of different modes are shown with horizontal lines and the position of the nodes of the corresponding eigenfunctions are represented with diamonds. 

In the hot and dynamically evolving environment of a PNS, mode classification becomes more challenging than in the case of cold NSs. To classify the non-radial modes of oscillation, we apply different  classification schemes that provide complementary physical information about the mode characteristics and insights into their evolution. The three different approaches we use are described in the following subsections.

\subsection{Cowling classification}
\label{subsec: cowling_class}

This classification scheme introduced by \cite{Cowling1941} is based on the restoring force (pressure or buoyancy),
acting on a fluid element when it is displaced from its equilibrium position inside the star. The number of nodes, \textit{n}, in the radial eigenfunction $\eta_r$, i.e. the number of times $\eta_r$ changes sign inside the star, is used to classify the modes. 
In this classification, the fundamental $f$-mode has $0$ nodes, and each $p_n$-mode and $g_n$-mode have $n$ nodes, with $n \geq 1$. The way to distinguish a $p$-mode from a $g$-mode with the same number of nodes is by comparing their frequencies: $p$-modes have higher frequencies than $g$-modes, and $f$-modes are in the middle. 

In the top panel of Figure~\ref{fig:class_methods}, we show as an example the radial eigenfunction profiles of the modes depicted in Figure \ref{fig:prop_diagram}. According to this classification, the green curve corresponds to a $p_1$-mode, with $1$ node in the $p$-propagation zone represented with a filled diamond. The yellow curve corresponds to a $g_2$-mode with $2$ nodes, one in the $p$-propagation zone and the other in the evanescent zone represented with filled and empty diamonds, respectively. Finally, a $g_1$-mode is represented with the purple curve, having $1$ node in the $g$-propagation zone, represented with an empty diamond. In this example, there is no $f$-mode between $p$-modes and $g$-modes. This classification is also used in \cite{Sotani2016,Torres2018,Morozova2018,Ferrari2003,Sotani2019,Sotani2020,Sotani2020b,Sotani2020c} for the modes of PNS, among others.  
Other works adopting this classification scheme for cold NS are, for example,  \citet{lindblom1983,mcdermott1988,benhar2004,lasky2015,Vasquez2017,Ranea2018} and \citet{Kumar_2023}. 

\subsection{Generalized Cowling Nomenclature}
\label{subsec:general_class}

\begin{figure}
    \centering
    \includegraphics[width=.97\linewidth]{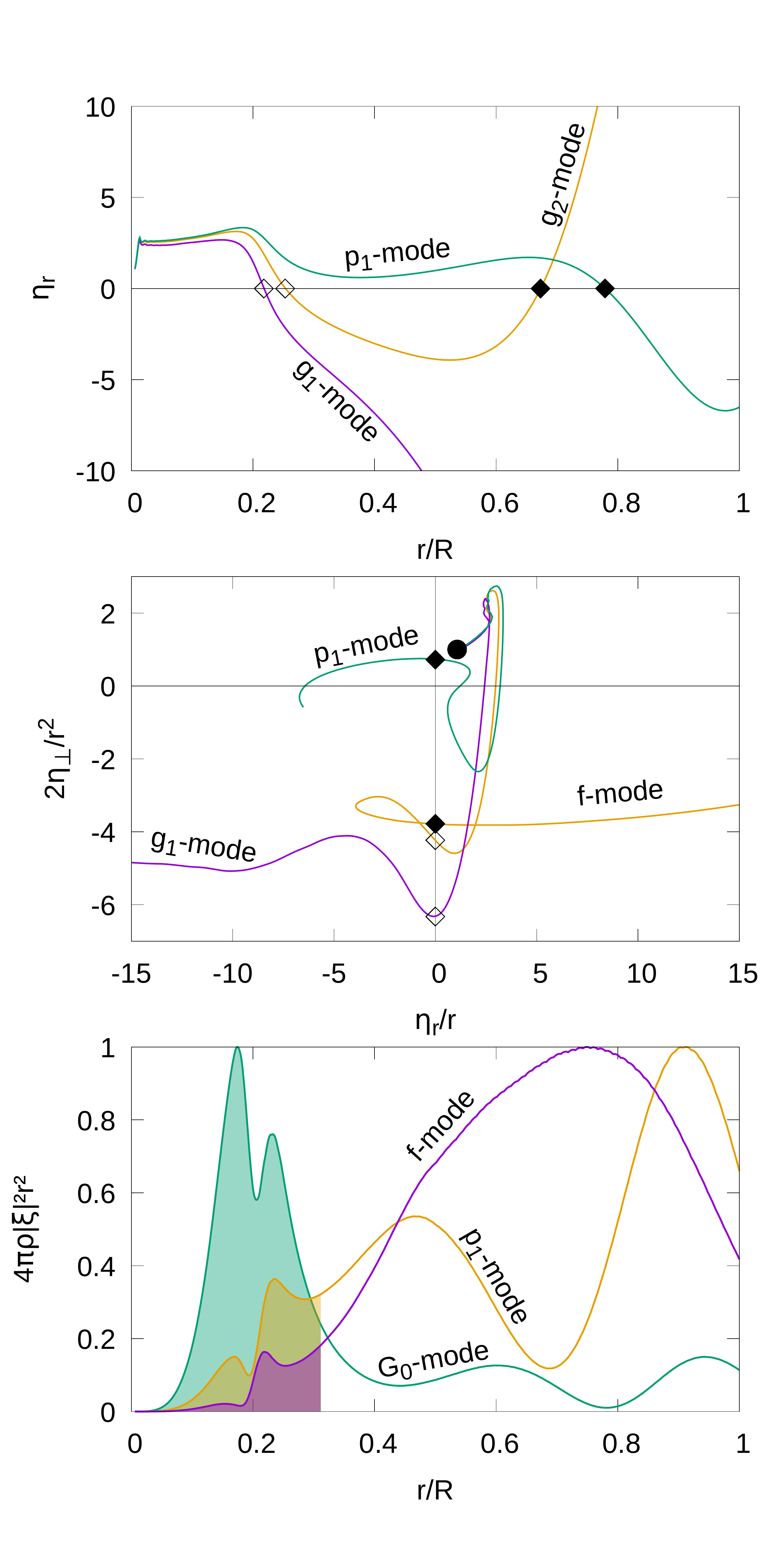}
    \caption{Representative example of the classification of PNS oscillation modes (see also the propagation diagram on Figure \ref{fig:prop_diagram}). From top to bottom, the mode identifications are obtained with the Cowling classification, the GCN and the CBMP, as described in the main text. The same three modes are represented in every panel with the same colors, but they are classified differently according to each method. On the top and middle panels, filled (empty) diamonds correspond to $p$-($g$-)nodes; on the bottom panel, the shaded range corresponds to the PNS core.}
    \label{fig:class_methods}
\end{figure}

The GCN classification \citep{Eckart1960,Scuflaire1974,Osaki1975}, takes into account the fact that, for evolved stars, the relation between the number of nodes in the eigenfunction and the mode number is not direct, in other words, modes do not necessarily have the same number of nodes along the stellar evolution, instead they can lose or gain nodes.  
The central panel of Figure~\ref{fig:class_methods} illustrates the phase diagram $(\eta_r,\eta_{\perp})$ for the modes depicted in the top panel with the same colours. The solid black circle indicates the stellar centre, and the nodes of the eigenfunctions are classified into \textit{g-nodes} and \textit{p-nodes}, depending on whether the curve in the phase diagram is travelling clockwise (g-nodes) or counterclockwise (p-nodes) at the crossing of the axis $\eta_r = 0$, as the radial coordinate, $r$, increases \citep{unno1989nonradial}. Accordingly, the empty diamonds represent g-nodes, and the filled diamonds represent p-nodes. 
The number of p-nodes and g-nodes are $N_p$ and $N_g$, respectively. Their difference  
\begin{eqnarray}
    \tilde{n}\equiv N_p-N_g ,
\end{eqnarray}
increases monotonically with the mode frequency, and it is conserved for each mode as the star evolves.
The modes are classified depending on the sign of $\tilde{n}$, which defines whether the mode is a $g$-mode ($\tilde{n}<0$) or a $p$-mode ($\tilde{n}>0$). For the fundamental mode, $\tilde{n}=0$. Hence, using this classification in the middle panel of Figure \ref{fig:class_methods}, the green curve shows a $p_1$-mode as it has $N_p=1$ and $N_g=0$ ($\tilde{n}=1$). The yellow curve belongs to the fundamental mode as it has $N_p=1$ and $N_g=1$ ($\tilde{n}=0$). Finally, the purple curve shows a $g_1$ mode as it has $N_p=0$ and $N_g=1$ ($\tilde{n}=-1$). Therefore, unlike in the Cowling Classification, there is a fundamental mode with two nodes between the $g$-modes and $p$-modes.

\subsection{Classification based on modal property}
\label{subsec:modal_class}

Even though the GCN is more appropriate to classify modes of evolved stars than the Cowling classification, it does not necessarily represent the character of the mode. Thereby, a second complementary classification, the CBMP, was introduced by \citet{Shibahashi1976}. This classification is based on the characteristics of the main trapping zone of the oscillation mode. In the propagation diagram, if the two propagation zones are widely separated by an intermediate evanescent zone, oscillation modes can be classified into two types by identifying their main trapping zones. In general, evolved main sequence stars can consist of four layers starting at the centre: a homogeneous convective core, a $\mu$-gradient zone with varying chemical potential $\mu$, a thin convective shell that acts like an evanescent zone and the radiative envelope, see e.g. Figure 1 of \citet{Shibahashi1976}. In analogy to the case of an evolved main sequence star, we use the CBMP to classify modes according to whether they are trapped in the core or the envelope of the PNS (see Figure \ref{fig:prop_diagram}, where the corresponding boundary between core and envelope is $r \sim 0.3R$). 
In order to describe this classification, some terminology needs to be defined
\begin{enumerate}
    \item $\textit{G}_n$: indicates a mode trapped mainly in the core zone that has \textit{n} $g$-nodes.
    \item $\textit{p}_n$: denotes a mode trapped mainly in the $p$- propagation zone of the envelope and having \textit{n} $p$-nodes.
    \item $\textit{g}_n$: a mode trapped mainly in the $g$-propagation zone of the envelope with \textit{n} $g$-nodes.
\end{enumerate} 
A key quantity for this classification scheme is the ratio of the kinetic energy of the oscillation mode within the core to that of the whole star, defined as:
\begin{eqnarray}
    \Delta=\frac{\int_{0}^{r_s}\rho |\xi|^2 4\pi r^2dr}{\int_{0}^{R}\rho |\xi|^2 4\pi r^2dr},
    \label{eqn:delta}
\end{eqnarray}
where $r_s$ is the outer radius of the core and 
the square of the displacement vector, $\xi^2$, is given by
\begin{eqnarray}
    \xi^2 = \eta_r^2 + l(l+1)\eta_\perp^2/r^2.
    \label{eqn:xi2}
\end{eqnarray}
According to this classification, if $\Delta$ is close to $0$ then the mode is trapped in the envelope. Otherwise, the mode is trapped in the core. The threshold, $\Delta_{\rm critic}$, used to define whether a mode is trapped in the core or in the envelope is an arbitrary parameter that we will discuss in section \ref{sec:results}. The number of nodes of each mode is determined by the number of nodes in the region where the mode is trapped. 

In the bottom panel of Figure~\ref{fig:class_methods}, we show the integrand in Eq.~(\ref{eqn:delta}) as a function of the radius of the PNS for the modes shown in the top and middle panels. In this example, the $p_1$-mode in the GCN (green curve) has $\Delta=0.66$, which means the mode is oscillating mostly in the core, without nodes in the core, so it is classified as a $G_0$-mode in this classification. On the other hand, the $f$-mode in the GCN (orange curve) has $\Delta=0.04$ and $1$ node in the $p$-propagation zone, so it is classified as a $p_1$-mode in this classification. Something similar occurs with the $g_1$-mode in the GCN (purple curve), which is classified as an $f$-mode because $\Delta=0.15$, and it has no nodes in the envelope. Unlike the other methods, the advantage of this classification is that it gives us information about the region of the star which contributes the most to the GW emission. 
%%%%%%%%%%%%%%%%%%%%%%%%%%%%%%%%%%%%%%%%%%%%%%%%%%%%%%%%%%%%%%%%%%%%%%%%%%%%%%%%
\section{Results}
\label{sec:results}

\subsection{PNS modes}
We consider only the quadrupolar ($l=2$) modes as they should be the most energetic in the gravitational-wave radiation from PNSs. 

To classify non-radial modes in PNSs, we consider $10$ models of PNS evolution, shown in Table \ref{tab:models}.  We calculate the frequencies of the modes in the Cowling approximation at every $10$ milliseconds in each of the models 
and we classify the modes using the three different classification methods described in section \ref{sec:classification}, focusing on the frequency range $100-2000$ Hz.

\begin{figure}
    \centering
    \includegraphics[width=.97\linewidth]{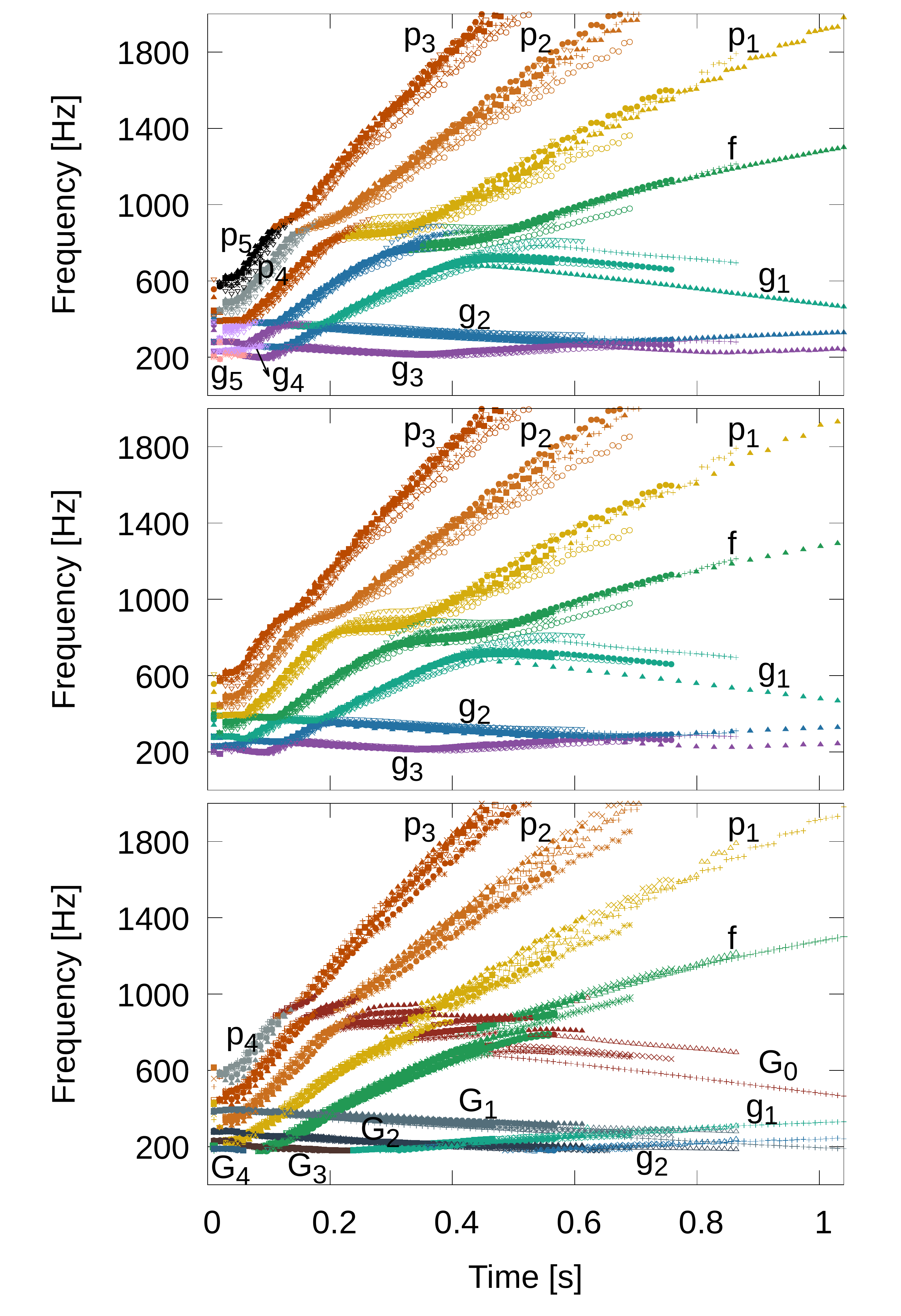}
    \caption{Time evolution of mode frequencies after the bounce for the PNS models shown in Table \ref{tab:models}. The same modes are shown in every panel, but from top to bottom the classifications used are the Cowling Classification, GCN and CBMP, respectively. For times greater than approximately 0.4 s after the bounce, all classifications agree on the identification of $f$- and $p$-modes. For earlier times, the Cowling classification is not well defined to the disappearance of the $f$-mode; the GCN shows several avoided crossings and the CBMP identifies an additional mode, $G_0$. See the main text for more details.}
    \label{fig:freq_time}
\end{figure}

In the Cowling classification, the modes are classified by counting the number of nodes of the amplitude of the radial Lagrangian displacement, $\eta_r$, and by comparing their frequency with that f the $f$-mode (with zero nodes) as discussed in section~\ref{subsec: cowling_class}. The results are presented in the top panel of Figure~\ref{fig:freq_time}. The different  symbols represent the PNS models introduced in Table \ref{tab:models} and the colours represent each mode classification, also indicated with labels in the figure. 
(There are modes with $4$ and $5$ nodes in other regions of the figure but we only represent those necessary to complete the mode evolution.) 
Strictly speaking, the Cowling approximation is only valid after $t \gtrsim 0.35$ s, when the $f$-mode is present.\footnote{We choose to follow \cite{Morozova2018} and classify the mode with 2 nodes at $0.1\,\textrm{s} \lesssim t \lesssim 0.4\, \textrm{s}$ as a $g_2$ mode, but the classification is not unique (it could also be classified as a $p_2$ mode) due to the absence of the $f$-mode at early times.} Going backwards in time from $t \sim 0.35$ s, we can see that modes gain nodes along the PNS evolution towards the bounce, when their frequencies get very close to each other, changing their Cowling classification. This feature is also seen in \cite{Morozova2018,Torres2019,Sotani2019}. Additionally, \cite{Sotani2019b,Sotani2020b} found that the amplitude of the modes involved in each ``close approach'' (see the avoided crossing discussion below) changes while the amplitude of the rest of the modes remains almost unchanged. 

The mode classification obtained with the GCN is presented in the middle panel of Figure~\ref{fig:freq_time}. 
The different colours now represent the classification through fixed values of $\tilde{n}=N_p-N_g$, which increases monotonically with the frequency at every time step. Progressing upward from the lowest frequency, the modes are uniquely classified as $g_3$, $g_2$, $g_1$, $f$, $p_{1}$, $p_{2}$ and $p_{3}$ (only modes with $\lvert \tilde{n} \rvert \leq 3$ are represented). Compared with the Cowling classification, in the GCN case we find that the description of the mode evolution is much simpler. The ``close approaches'' noted in the top panel now appear as avoided crossings. 
For instance, as the $f$-mode evolves, it shows avoided crossings as it bounces between the $p_1$- and $g_1$-modes: at $t \simeq 0.05$ s with $p_1$, at $t \simeq 0.15$ with $g_1$, at $t \simeq 0.3$ s with $p_1$ and finally at $t \simeq 0.5$ s with $g_1$.

The results obtained in the CBMP are presented in the bottom panel of Figure~\ref{fig:freq_time}. The colours represent different mode classifications. Modes with upper case($G_i$-modes) and lower case ($p_i$, $f$ and $g_i$-modes) letters correspond to modes trapped in the core and in the envelope, respectively. The sub-index in each label represents the number of nodes in the region where the mode is trapped. For instance, the $G_2$-mode is trapped in the core and has $2$ nodes in the core, the $f$-mode is trapped in the envelope and has no nodes in the envelope (but it could have nodes in the core), and the $p_2$-mode is trapped in the envelope and has $2$ nodes in the $p$-propagation zone in the envelope. We note that the $G_i$-modes decrease in frequency over time, and there are no more avoided crossings. In particular, the $G_0$-mode appears, crossing all $p$-modes and the $f$-mode.

It is remarkable that all of the PNS models we analyze show very similar results. 
\begin{figure}
    \centering
    \includegraphics[width=.97\linewidth]{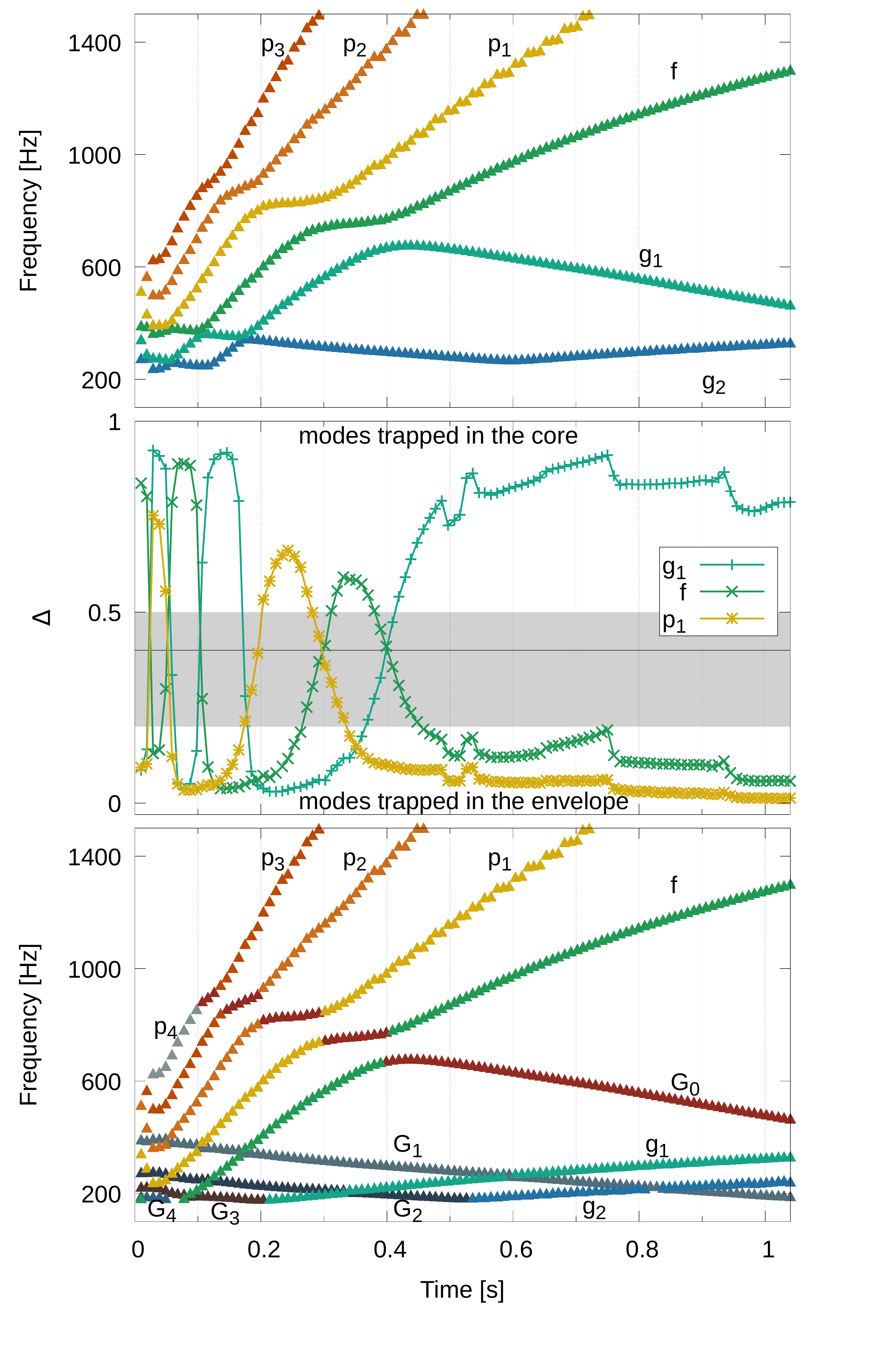}  \caption{Top: time evolution of mode frequencies classified with the GCN for PNS model 1 in Table \ref{tab:models}. Middle: time evolution of the ratio between the kinetic energy of the oscillations within the core and their total kinetic energy inside the star, for the $p_1$, $f$, and $g_1$-modes in the top panel. The horizontal line shows the threshold in $\Delta$ used for the CBMP, and the shaded region represents  possible thresholds of $\Delta$ that give qualitatively the same result (see also Appendix \ref{Appendix:C}). Bottom: time evolution of the mode frequencies for the same model shown in the top panel, now classified using the CBMP.}
    \label{fig:delta_freq}
\end{figure}
As a representative example, we choose model 1 (with a $9 M_{\odot}$ progenitor star) to explore the complementary descriptions provided by the GCN and CBMP. In Figure ~\ref{fig:delta_freq} we present the mode evolution only for model 1, using the GCN and the CBMP in the top and bottom panels, respectively. In the central panel, we show the evolution of $\Delta$ for the $g_1$-, $f$- and $p_1$-modes, as classified by the GCN. The horizontal black line corresponds to the $\Delta$ threshold we use in this work. The grey band represents a $\Delta$ range for which the classification would be qualitatively the same (see more details in Appendix \ref{Appendix:C}). Values of $\Delta$ above the horizontal black line indicate the mode is trapped in the core. Otherwise, the mode is trapped in the envelope of the PNS. Near each avoided crossing in the GCN, $\Delta$ changes abruptly and the character of the mode changes (trapped in the core or trapped in the envelope).

\begin{figure*}
    \centering
    \includegraphics[width=.8\linewidth]{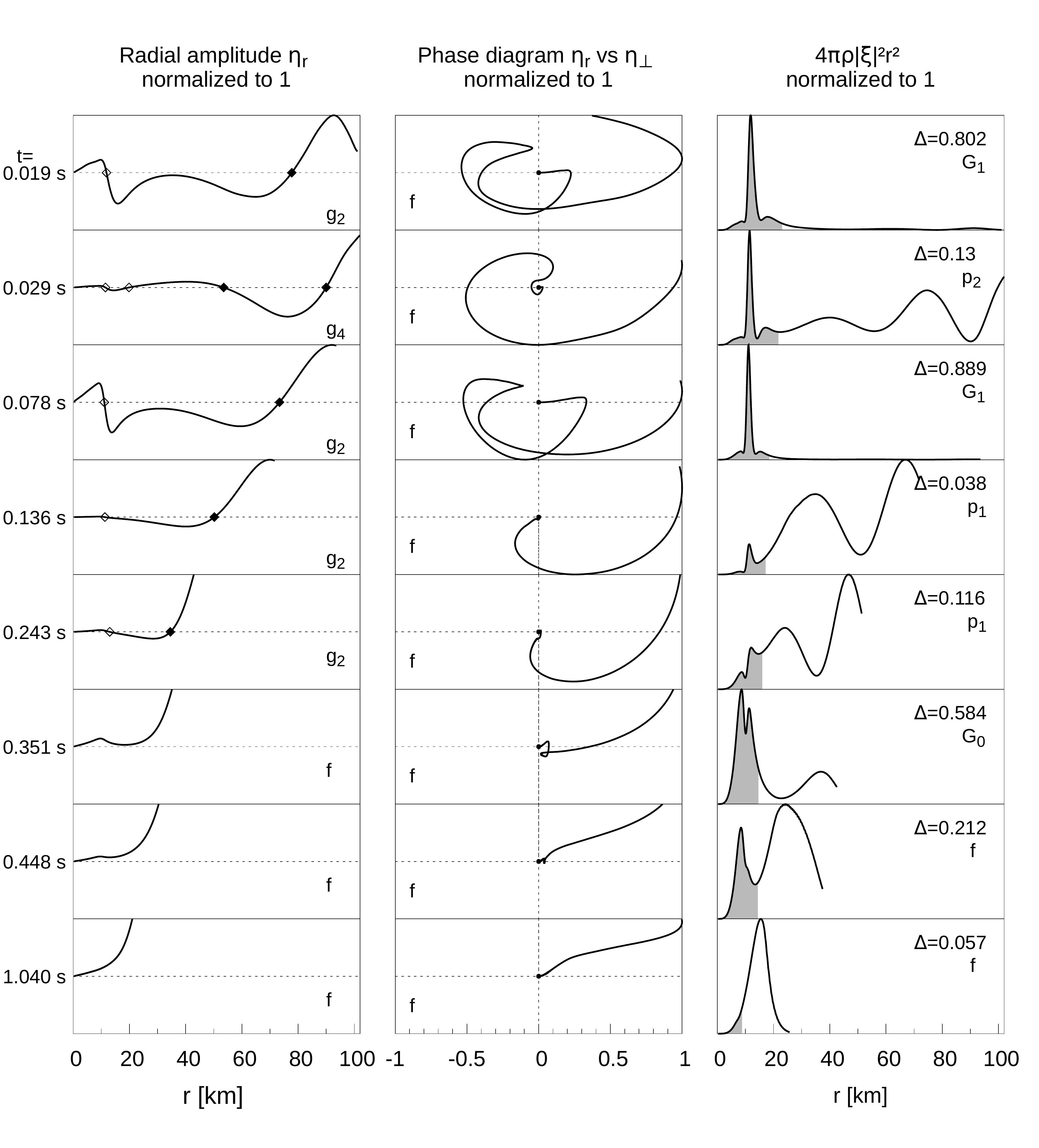}  \caption{Evolution of the parameters used in each mode classification method, corresponding to the $f$-mode (as classified by the GCN). Times are specified on the left side and mode classifications are shown as labels in each panel. Left: evolution of the radial eigenfunction $\eta_r$, used in the Cowling classification method. Center: mode trajectories in the $(\eta_r,\eta_{\perp})$ phase space, used in the GCN. Right: evolution of $\Delta$ as a function of the radius of the PNS. 
    }
    \label{fig:class_panels}
\end{figure*}

We take a closer look at the evolution of the $f$-mode (as classified by the GCN) in Figure~\ref{fig:class_panels}. For 8 time slices, we present the radial eigenfunction in the left columns, the $(\eta_r,\eta_{\perp})$ phase space in the center and the integrand used in the definition of $\Delta$ in the right column (note that the PNS becomes more compact with time, as it cools and approaches the cold NS stage). The $f$-mode gains and loses pairs of $p$- and $g$-nodes, changing its Cowling classification but keeping $\tilde{n}$ constant in the GCN. The CBMP shows the changes in the mode character, alternating between oscillations trapped in the core ($G_i$ classifications) and in the envelope ($p_i$ and $f$ classifications).

\subsection{Emission of gravitational waves}

\begin{figure}
    \centering
    \includegraphics[width=1.0\linewidth]{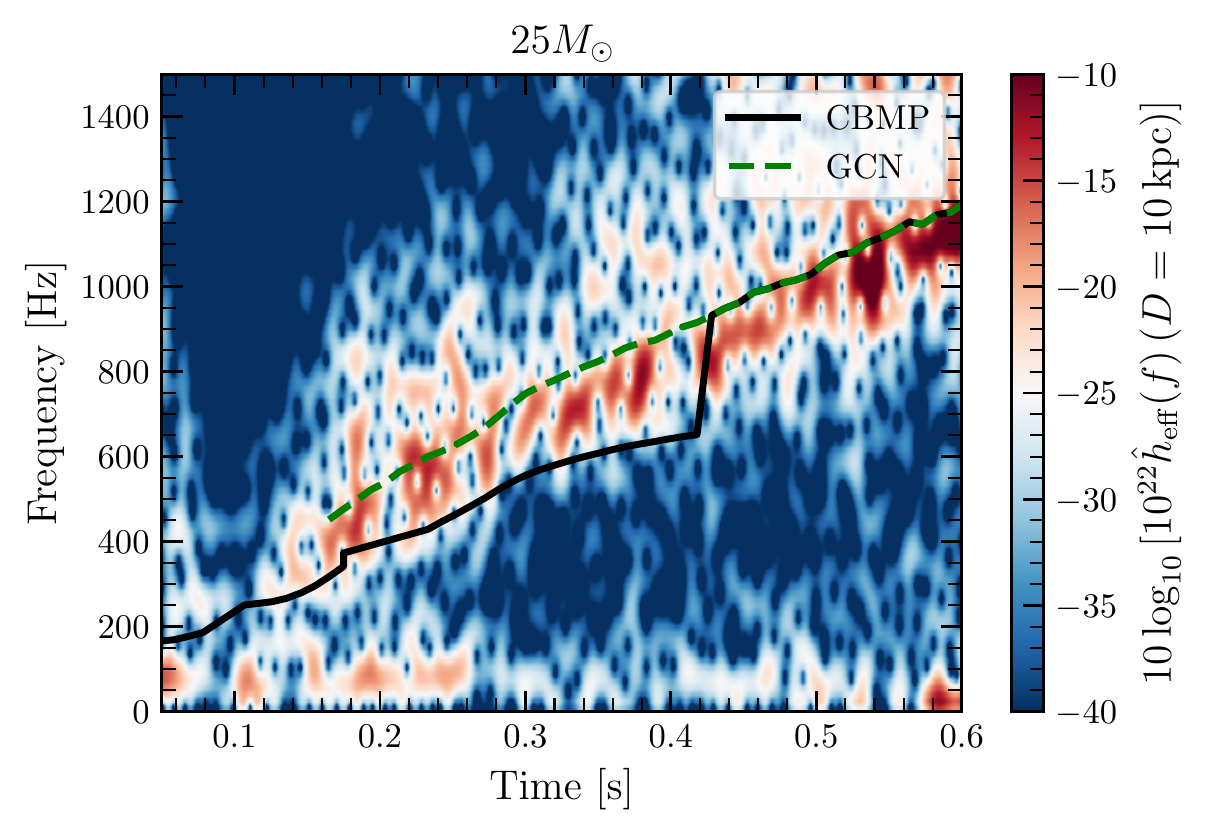}  \caption{Spectrogram of the GW emission for model 9 (see Table \ref{tab:models}), shown together with the $f$-mode sequences identified with the GCN and the CBMP. For earlier times, the GCN shows better agreement with the simulated GWs. For later times, both methods agree on the $f$-mode classification. }
    \label{fig:spec}
\end{figure}

It is relevant to ask how most of the GW emission by the PNS is generated. A comparison between the time-evolving spectrum of modes and the spectrogram of the GWs for model 9 is presented in Figure \ref{fig:spec}.

We show with lines the evolution of the $f$-mode obtained following the GCN classification and the CBMP. In this analysis, we use the modes calculated with the full eqs.~(\ref{eqn:1})-(\ref{eqn:4}) in order to obtain a better agreement with the simulated GWs. As the GCN and CBMP are strictly valid only in the Cowling approximation of eqs.~(\ref{eqn:5})-(\ref{eqn:6}), the classifications shown here were obtained in comparison with the results for the Cowling approximation in the same model.
The GCN $f$-mode shows very good agreement with the strongest GW emission and not very prominent avoided crossings at least at $t \simeq 0.15$ and \mbox{0.4 s}. This result, together with the detailed analysis of the GCN $f$-mode presented in Figure \ref{fig:class_panels}, indicates that the PNS spectrum evolves with avoided crossings. Consequently, it emits GWs mainly through an $f$-mode that rapidly alternates between being trapped in the core and in the envelope of the PNS. We note that, despite the discrepancy between the CBMP $f$-mode and the gravitational wave emission at early times, the CBMP classification is still applicable and can provide interesting information, for example helping in the characterization of the modal properties of the GCN $f$-mode  and in the identification of avoided crossings.

\subsection{Quasi-universal relations}

Universal relations between mode frequencies and macroscopic parameters, such as mass and radius, have been proposed as a possible tool to infer properties of NSs using the detection of GWs, see e.g.  \citet{Anderson:1998tgw,benhar2004,Chirenti2015,tsui:2005uiq,lau:2010ipp,sotani:2021urb,ranea-sandoval:2023cmr}. 

Here we analyze some quasi-universal relations proposed in the literature for cold NSs in the context of the dynamic PNS. As all of our PNS models have been produced with the SFHo EoS for dense matter derived by \citet{Steiner2013}, we are only able to probe the influence of the progenitor mass. 

It has been shown that the frequency of the fundamental mode of cold NSs scales with $\sqrt{M/R^3}$ with minor dependence on the EoS \citep[see, for example,][]{Anderson:1998tgw,benhar2004,Chirenti2015}.  \cite{Torres2019b} proposed a similar dependence for the frequencies of $f$ and $p$-modes of a PNS, see also \citet{Sotani2021} and \citet{Mori2023}. 

However, note that there is no unique definition for the radius of PNS. \cite{Torres2019b} considers that the PNS extends up to the radial position of the shock wave, while we follow \cite{Morozova2018} and use $\rho = 10^{10}$ g/cm$^{3}$ as a condition to define the radius, which gives somewhat larger values (but both choices produce similar values for $\sqrt{M/R^3}$).

\begin{figure}
    \centering
    \includegraphics[width=1.0\linewidth]{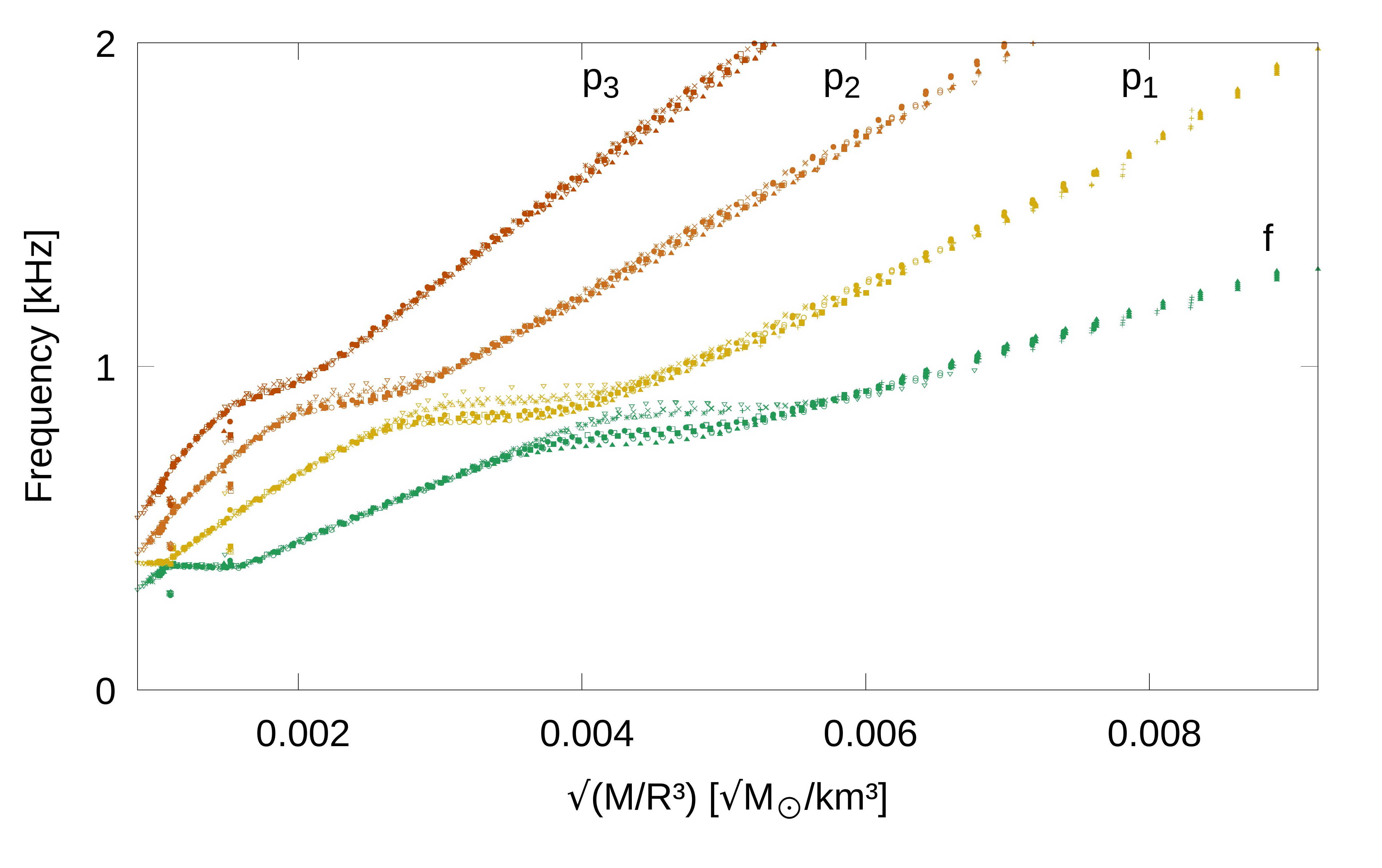} 
        \includegraphics[width=1.0\linewidth]{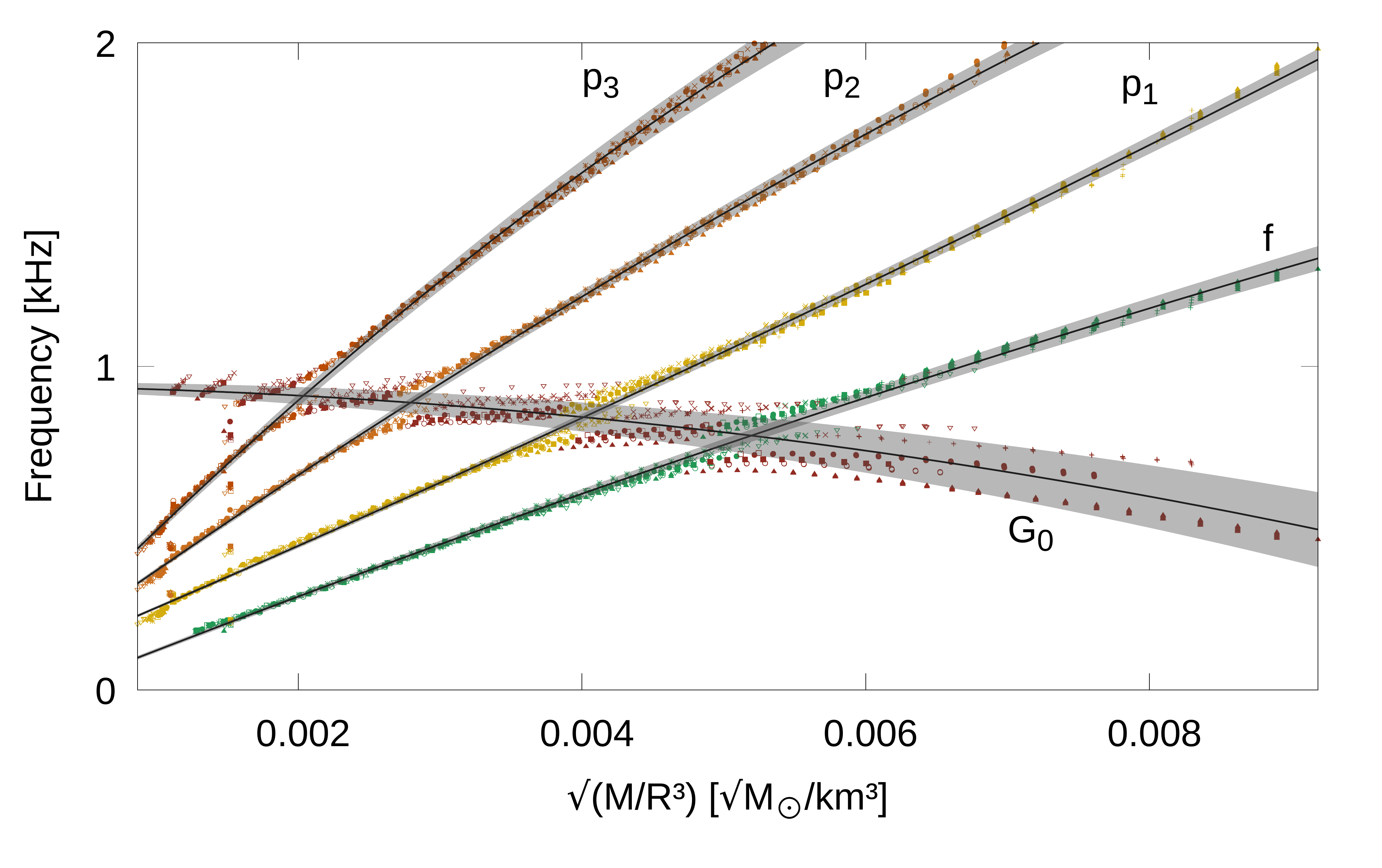} 
    \caption{Relation between macroscopic quantities (mass, $M$, and radius, $R$) and the frequencies of modes classified using the GCN (top panel) and the CBMP (bottom panel). 
    The CBMP sequences show simpler trends than the GCN sequences, due to the absence of the avoided crossings. 
    The black lines represent the fits to each mode in the CBMP (see Table \ref{tab:fits}) and the shaded areas represent the $\pm 1\sigma$ bands of each fit. Note that the frequencies corresponding to the $G_0$-modes below the $f$-modes in the bottom panel are not included in the top panel because they are classified as $g_1$-modes in the GCN.}
    \label{fig:universal-relations}
\end{figure}

In the top panel of figure~\ref{fig:universal-relations} we show sequences of modes classified with the GCN method. As previously, the symbols represent the different models, and each colour corresponds to a mode.  
The avoided crossings spread the curves, which are mostly independent of the progenitor mass (time increases to the right).
In the bottom panel of figure~\ref{fig:universal-relations}, we present the sequences of modes classified with the CBMP.  
Now $f$ and $p$-modes present a very simple monotonic behaviour which is independent of the progenitor mass. The $G_0$-mode also shows a simple trend but with a larger spread. The fits to this novel quasi-universal relation for the modes classified with the CBMP is presented in Table \ref{tab:fits}. 

\begin{center}
\label{tab:fits}
\begin{table}
\begin{tabular}{||c c c c c||} 
 \hline
 Mode & a [Hz] & b [${\rm Hz}\, {\rm km}^{3/2}/M_{\odot}^{1/2}$] & c [${\rm Hz}\, {\rm km}^{3}/M_{\odot}$]\\ [0.5ex] \hline\hline %& final sum of squares of residuals \\ [0.5ex]  \hline\hline
 $f$ & $-0.050\pm0.004$ & $174.3\pm1.9$ & $-2590\pm195$ \\ %& $0.226406$ \\ 
 \hline
 $p_1$ & $0.067\pm 0.003$ & $184.6\pm1.6$ & $2194\pm 176$ \\ %& $0.311872$ \\
 \hline
 $p_2$ &  $0.062\pm 0.004$ & $313.2\pm2.5$ & $-6213\pm326$ \\ %& $0.329435$ \\
 \hline
 $p_3$ & $0.057\pm 0.006$ & $452.6\pm4.9$ & $-16780\pm807$ \\ %& $0.364522$ \\
 \hline
 $G_0$ & $0.94\pm 0.01$ & $-4.8\pm5.7$ & $-4715\pm596$ \\ [1ex]%& $1.9904$ \\ [1ex] 
 \hline
 %$g_1$ & $M/R^2$ & $0.137\pm0.001$ & $87.199\pm1.297$ & $-$ \\ [1ex]%& $0.0482453$ \\ [1ex] 
 %\hline
 %$g_2$ & $M/R^2$ & $0.118\pm0.004$ & $55.19\pm2.42$ & $-$ \\ [1ex]%& $0.00921777$ \\ [1ex] 
 %\hline
\end{tabular}
\caption{Best fit parameters for quasi-universal relations of the form \mbox{$f(x)=a+bx+cx^2$}, describing the CBMP mode frequency $f$ in Hz as a function of $x \equiv \sqrt{M/R^3}$ in units of $M_{\odot}^{1/2}/{\rm km}^{3/2}$, shown in Figure \ref{fig:universal-relations}. 
}
\end{table}
\end{center}

%%%%%%%%%%%%%%%%%%%%%%%%%%%%%%%%%%%%%%%%%%%%%%%%%%%%%%%%%%%%%%%%%%%%%%%%%%%%%%%
\section{Conclusions and Discussion}
\label{sec:conclusions}

The asteroseismology of PNSs is complicated by their dynamic and temperature-dependent nature. In this paper, we consider three different methods to classify their non-radial modes of oscillation: the Cowling classification, the GCN and the CBMP.

The standard Cowling classification is not well defined for a PNS at early times after the bounce. It relies on the identification of a fundamental $f$-mode without nodes to distinguish between $p$-modes (with higher frequencies) and $g-$modes (with lower frequencies). This mode is not present at early times, causing the mode identifications to be underdetermined.

The GCN relies on the identification of $p$- and $g$-nodes and allows us to follow the mode evolution. It shows marked avoided crossing at early times, at which two modes exchange character. The $f$-mode identified with the GCN has very good agreement with the simulated GW emission of a PNS. Interestingly, this implies that the region responsible for most of the GWs (where the mode is trapped), quickly alternates between the core and envelope of the star at early times.

The CBMP follows the time evolution of sequences of modes with the same properties (trapped in the core or trapped in the envelope) and it is applied here for the first time to the study of PNS oscillation modes. The behavior of these sequences is remarkably simple and monotonic, as the avoided crossings are absent, and we present fits for quasi-universal relations (independent of progenitor mass) describing the most relevant mode frequencies as functions of the PNS average density.  We note that the matching classification scheme 
 of \cite{Torres2019}, in which modes are classified according to the similarities between their eigenfunctions, produces results qualitatively similar to the CBMP, showing no avoided crossings at early times and a mode compatible with the $G_0$-mode, classified there as a $g_2$-mode. Additionally, the ramp-up mode discussed by \citet{Mori2023} is similar to the $f$-mode described by the CBMP. One of the main advantages of the CBMP is that it gives physical information relative to the part of the PNS where GWs are most likely being produced.

There are a few different promising directions for extensions of our work. First, formal work should be done to extend the GCN and CBMP criteria beyond the Cowling approximation ($\delta \alpha \neq 0$) \footnote{See also an alternative approach by \citet{Takata2012} for taking the perturbation of the gravitational potential into account by using a continuous integral index.} and to calculate the modes using time-dependent perturbation theory. Second, the quasi-universal behavior of the mode frequencies should be checked against simulations with different progenitor masses and additional EoSs. 

Finally, an assessment of the detectability (and reliable identification) of these modes in a galactic core-collapse supernova would be of great interest. Recently, \citet{Bruel2023} have estimated that current GW detectors could infer properties of such events up to the Large Magellanic Cloud. The mode evolution provides the rate of contraction of the PNS, which depends not only on the EOS but also on transport properties in dense matter, related to the rate of deleptonization of the star. Abundant neutrinos detected from a galactic core-collapse supernova will provide complementary information \citep{Nagakura:2021yci}, potentially allowing for the independent determination of the time-evolving mass and radius of the PNS.

%%%%%%%%%%%%%%%%%%%%%%%%%%%%%%%%%%%%%%%%%%%%%%%%%%%%%%%%%%%%%%%%%%%%%%%%%%%%%%%
\section*{Acknowledgements}
The authors gratefully acknowledge Viktoriya Morozova for guidance related to the PNS models and QNM calculation during the early stages of this project, and Adam Burrows and Shin Yoshida for useful comments and discussions.

M.C.R.~and I.F.R.-S.~acknowledge CONICET and UNLP for financial support under grants G157 and G007. I.F.R.-S.~is also partially supported by PICT 2019-0366 from ANPCyT (Argentina) and by the National Science Foundation (USA) under Grant No. PHY-2012152. 
C.C.~acknowledges support by NASA under
award numbers 80GSFC17M0002 and TCAN80NSSC18K1488.
D.R.~acknowledges funding from the U.S. Department of Energy, Office of Science, Division of Nuclear Physics under Award Number(s) DE-SC0021177 and from the National Science Foundation under Grants No. PHY-2011725, PHY-2020275, PHY-2116686, and AST-2108467.

%%%%%%%%%%%%%%%%%%%%%%%%%%%%%%%%%%%%%%%%%%%%%%%%%%%%%%%%%%%%%%%%%%%%%%%%%%%%%%%
\section*{Data Availability}
 
The computed data presented and discussed in this paper will be shared upon reasonable request.


\bibliographystyle{mnras}
\bibliography{bibliography} % if your bibtex file is called bibliography.bib

%\begin{thebibliography}{99}
%\end{thebibliography}

%%%%%%%%%%%%%%%%%%%%%%%%%%%%%%%%%%%%%%%%%%%%%%%%%%
%%%%%%%%%%%%%%%%% APPENDICES %%%%%%%%%%%%%%%%%%%%%
\onecolumn
\appendix

%%%%%%%%%%%%%%%%%%%%%%%%%%%%%%%%%%%%%%%%%%%%%%%%%%
%\newpage
\section{Results for different \texorpdfstring{$\Delta$}{TEXT} thresholds} \label{Appendix:C}

In the CBMP, modes are classified by the region in the star where they are trapped. With this method, the modes change character when the ratio of the kinetic energy  of the oscillations within the core to that of the whole star, $\Delta$, significantly changes from $\sim 0$ to $\sim 1$, and vice versa. 
In this appendix, we test different values of $\Delta_{\rm{critic}}$ from which we determine if the mode is trapped in the core or in the envelope of the PNS.

\begin{figure*}
    \centering
    \includegraphics[width=.8\linewidth]{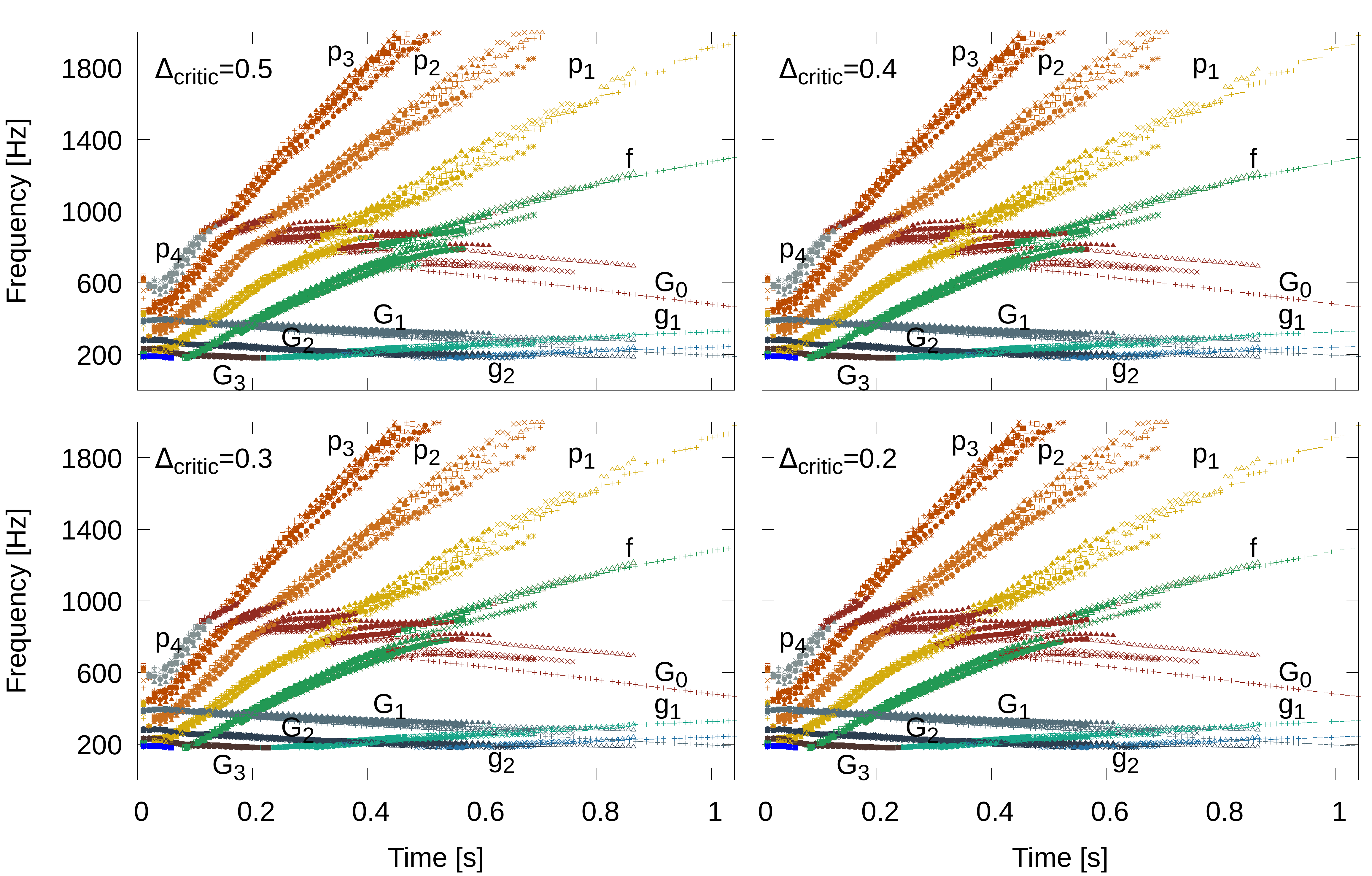}
    \caption{Time evolution of mode frequencies after the bounce using the CMBP for values of $\Delta_{\rm{critic}}$ between $0.2$ and $0.5$ (see middle panel of Figure \ref{fig:delta_freq}).  %indicated with labels in the top corner of each panel. The colors are associated with different modes shown with labels.
    The results are quantitatively the same in all cases; the core-trapped $G_0$-mode becomes slightly more prominent with decreasing $\Delta_{\rm{critic}}$. }
    \label{fig:different_deltas}
\end{figure*}

In figure \ref{fig:different_deltas}, we show the time evolution of the mode frequencies classified with the CBMP. In each panel, we used a different value of $\Delta_{\rm{critic}}$ from $0.2$ to $0.5$. 
The modes change character at different times for each case: 
for larger $\Delta_{\rm{critic}}$, modes trapped in the core (envelope) become trapped in the envelope (core) earlier (later) compared with smaller $\Delta_{\rm{critic}}$. 
Therefore, for $0.2 \leq \Delta_{\rm{critic}} \leq 0.5$, the results are qualitatively the same.
%%%%%%%%%%%%%%%%%%%%%%%%%%%%%%%%%%%%%%%%%%%%%%%%%%

% Don't change these lines
\bsp	% typesetting comment
\label{lastpage}
\end{document}